# Vorticity dynamics of revolving wings: The role of planetary vortex tilting on the stability of the leading-edge vortex

Nathaniel H. Werner[1], Hojae Chung[2], Junshi Wang[3], Geng Liu[3,4], John Cimbala[1], Haibo Dong[3], and Bo Cheng[1,†]

1. Department of Mechanical and Nuclear Engineering, Pennsylvania State University, University Park, PA 16802, USA
2. Department of Aerospace Engineering, Pennsylvania State University, University Park, PA 16802, USA
3. Department of Mechanical and Aerospace Engineering, University of Virginia, Charlottesville, VA 22904, USA
4. Department of Mechanical Engineering, University of Maine, Orono, ME 04469, USA

This work investigated the vorticity dynamics and stability of leading-edge vortices (LEVs) in revolving wings. Previous studies suggested that Coriolis acceleration and spanwise flow both played key roles in stabilizing the LEV; however, the exact mechanism remains unclear. The current study examined a mechanism that relates the effects of Coriolis acceleration, spanwise flow, and the tilting of the planetary vortex on limiting the growth of the LEV. Specifically, this mechanism states that a vertical gradient in spanwise flow can create a vertical gradient in Coriolis acceleration, which will in turn produce oppositely-signed vorticity within the LEV. This gradient of Coriolis acceleration corresponds to the spanwise (radial) component of planetary vortex tilting (PVTr) that reorients the planetary vortex into the spanwise direction therefore creating oppositely-signed LEV vorticity. Using an in-house, immersed-boundary-method flow solver, this mechanism was investigated alongside the other vorticity dynamics for revolving wings of varying aspect ratio ($AR = 3, 5$, and $7$) and Reynolds number ($Re = 110$, and $1400$). Analyses of vorticity dynamics showed that the PVTr consistently produced oppositely-signed vorticity for all values of $AR$ and Re investigated, although other three-dimensional phenomena play a similar but more dominant role when $Re = 1400$. In addition, the relative strength of the PVTr increased with increasing $AR$ due to a decrease in the magnitude of advection. Finally, the effects of $AR$ and $Re$ on the vorticity dynamics and LEV stability were also investigated.

## 1. Introduction

One of the most critical aerodynamic mechanisms that enable insect flight is the lift augmentation from leading-edge vortices (LEVs), which also exist in revolving wings operating at high angles of attack (Usherwood & Ellington 2002; Poelma *et al.* 2006; Lentink *et al.* 2009). Despite the variations in their integrity (Lentink & Dickinson 2009b; Jones *et al.* 2016) and secondary structures (Lu *et al.* 2006, 2007; Lu & Shen 2008; Harbig *et al.* 2013b; Garmann *et al.* 2013; Garmann & Visbal 2014), the primary LEV structure on insect wings does not grow into instability and remains attached to the wing (Lentink & Dickinson 2009a,b; Harbig *et al.* 2013b). Notably, the stability and attachment of the vortex structure are sustained not only transiently during the revolving period of the flapping wings (delayed stall), but also for wings undergoing continuous unidirectional rotation. These flow phenomena, presumably manifesting from the three-dimensional flow effects unique to revolving wings, are fundamentally different from those generated by their translating counterparts, which normally stall at high angles of attack.

†Email address: buc10@psu.edu

Recently, a number of studies have carefully examined the vortex structure and the corresponding vortex dynamics of revolving wings (Lentink & Dickinson 2009a,b; Kim & Gharib 2010; Cheng *et al.* 2013; Harbig *et al.* 2013b; Garmann & Visbal 2014; Wolfinger & Rockwell 2014; Carr *et al.* 2015; Jardin & David 2015; Jardin 2017; Smith *et al.* 2017) often with the goal of testing one of the four hypotheses related to LEV stability.

Arguably, the most well-known hypothesis has been proposed by Ellington *et al.* (1996) and has been further investigated by Birch & Dickinson (2001). It states that the spanwise flow in the region above the wing transports the LEV vorticity towards the tip, which is then shed within a coherent structure connecting the LEV and tip vortex (TiV). According to this hypothesis, the vorticity transport caused by spanwise flow is the main contributor to balancing the vorticity transported from the leading-edge. However, several studies that analyse the vorticity transport using either experiments (Cheng *et al.* 2013; Wojcik & Buchholz 2014) or simulations (Shyy & Liu 2007; Aono *et al.* 2008) question this hypothesis; as they find no direct evidence of significant vorticity transport caused by spanwise flow. Nonetheless, Garmann & Visbal (2014) observe a reduced degree of LEV stability caused by weakened spanwise flow. Additionally, Jardin & David (2014) by comparing flows with externally added spanwise gradients and flows over revolving wings, observe that a spanwise gradient in flow speed, viscous effects, and Coriolis and centrifugal acceleration all contribute to LEV attachment. However, they conclude that ultimately the Coriolis and centrifugal effects play the primary role in lift generation.

The second hypothesis put forth by Lentink & Dickinson (2009b) relates the stability of the LEV to the effects of Coriolis and centrifugal accelerations, arguing that both play a vital role in LEV attachment and are key mechanisms in the generation of additional lift (Jardin & David 2015; Jardin 2017). This hypothesis is based on observations of the distribution of wing aspect ratio in natural flyers (Chin & Lentink 2016; Lentink & Dickinson 2009a) as well as their investigations of Navier-Stokes (N-S) equations in a relative rotating frame and experiments using dynamically-scaled robotic wings (Lentink & Dickinson 2009b; Kruyt *et al.* 2015). Both Coriolis and centrifugal accelerations are inversely-proportional to Rossby number ($Ro$) (Lentink & Dickinson 2009a), which is defined as the ratio of the advective and Coriolis accelerations (Kundu *et al.* 2008). According to Lentink & Dickinson (2009b), wings in nature on average have wing tip $Ro$ between 3 and 4 (Chin & Lentink 2016; Lentink & Dickinson 2009a), leading them to conclude that sufficiently low $Ro$ and therefore a high Coriolis acceleration is necessary for LEV stability. It is important to note that in Lentink & Dickinson (2009b), the $Ro$ is equivalent to the $AR$ used here. Jardin & David (2015) further show that artificially removing the Coriolis acceleration reduces both the stability of the LEV and the lift because the Coriolis acceleration plays a role in maintaining the proximity of the LEV to the wing surface. Jardin (2017) observes at $Re > 200$ that Coriolis acceleration is the primary contributor to LEV attachment, while at $Re < 200$ the viscous effects dominate, and that centrifugal effects were negligible. Nonetheless, although the $Ro$ and therefore the relative magnitude of Coriolis acceleration are strong indicators of LEV stability, the Coriolis acceleration *per se*, as pointed out by Garmann & Visbal (2014), is unlikely to be a direct contributor to LEV stability since it points away from the wing, which may result in LEV detachment. Therefore, the exact mechanism of Coriolis acceleration in stabilizing the LEV, if it exists, remains unclear.

Finally, the two remaining hypotheses argue that either vortex annihilation due to interactions with a shear layer that develops between the LEV and wing surface (Wojcik & Buchholz 2014; Panah *et al.* 2015; Akkala & Buchholz 2017; Onoue & Breuer 2017) or vorticity transport caused by downwash being generated by the tip vortex (Birch & Dickinson 2001; Lauder 2001; Ozen & Rockwell 2012; Carr *et al.* 2013) are the primary source of LEV stability. Onoue & Breuer (2017) also conclude that vorticity annihilation is a dominant mechanism in regulating the growth of the LEV more so than the spanwise advection for a sweeping flat plate at $Re \approx \mathcal{O}(10^5)$. Wojcik & Buchholz (2014) and Panah *et al.* (2015) observe that the flux of oppositely signed vorticity generated at the wing surface is an important factor in controlling the strength of the LEV. In addition, Cheng *et al.* (2013) shows that there exists a significant downward advection of radial vorticity by downwash, which plays a key role in the vorticity dynamics.

This work establishes and tests a previously undiscussed mechanism that combines the roles of the Coriolis acceleration and the spanwise flow in contributing to LEV stability. It provides a new perspective towards the roles of spanwise flow instead of its vorticity transport, which is found negligible in previous studies (see above). Specifically, this mechanism states that the vertical gradient in spanwise flow (blue, figure 1) leads to a vertical gradient in the streamwise Coriolis acceleration (green, figure 1), which results in angular acceleration that is oriented opposite to the rotation of the LEV vorticity (purple, figure 1). Therefore, it creates oppositely-signed LEV vorticity and contributes to its stability. Previous experimental work by Cheng *et al.* (2013) shows a clear negative vertical gradient of the spanwise velocity field in the region of the LEV (i.e., the magnitude of the radial velocity increases with decreasing height above the wing), which has motivated the authors to test this mechanism. In the vorticity equation, which is the curl of the N-S equations (Batchelor 2000; Kundu *et al.* 2008), this mechanism is equivalent to the tilting of the planetary vortex into the outboard (root to tip) radial direction due to the vertical gradient of the spanwise velocity. Note that the planetary vortex is tangent to the rotation axis of the wing and exists everywhere in the relative rotating frame. This tilting of the planetary vortex, which we name radial planetary vortex tilting (PVTr), can be shown to be equal to the curl of the Coriolis acceleration. Therefore, the PVTr introduces oppositely-signed vorticity to the LEV, and contributes to preventing the LEV from growing unstable and shedding from the leading-edge. The PVTr is also expected to act opposite to the LEV vorticity regardless of the wing $AR$ or $Re$, as long as a vertical gradient of spanwise flow exists in the region of the LEV. Therefore, in this mechanism, it is the curl (or vertical gradient) of Coriolis acceleration due to the gradient of spanwise flow that contributes to the stability the LEV, instead of the Coriolis acceleration or spanwise flow themselves. In this work, based on the velocity and vorticity data obtained using an in-house immersed-boundary-method flow solver we examine this mechanism by considering the role of PVTr in the vorticity dynamics of revolving wings of different $AR$ (3, 5, and 7) and $Re$ (110, and 1400).

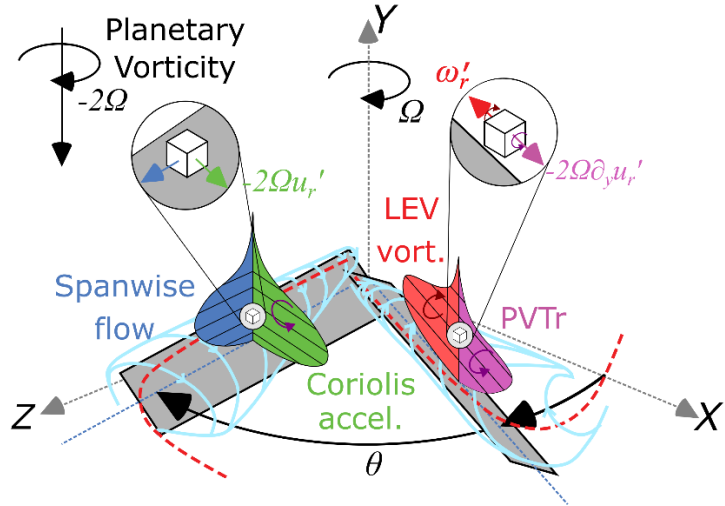

**Figure 1**. A schematic showing the mechanism of LEV stability based on the radial Planetary Vortex Tilting (PVTr). On the left wing, the spanwise flow gradient and streamwise Coriolis acceleration are drawn as hypothesized. On the right wing, the spanwise vorticity and PVTr are also drawn. A vertical arrow is also given indicating the sign and magnitude of the planetary vortex.

## 2. Materials and Methods

### 2.1 *Numerical Methods*

This work simulated six cases of revolving wings where each wing starts impulsively and then rotates at a constant angular velocity Ω for three full revolutions about a vertical axis aligned with the wing root at the mid-chord location (figure 2a). The rotation angle measured from the initial position is denoted by $\theta$. All wings have rectangular shape with a constant chord length c = 1 cm, angle of attack $\alpha = 45°$, and infinitesimal thickness. The six cases were defined according to three aspect ratios $AR = s/c$ of 3, 5, and 7 and two Reynolds numbers $Re = \Omega r_g c/\nu$ of 110 and 1400; where s is the wingspan length, $r_g$ is wing radius of gyration, and $\nu$ is the kinematic viscosity of the fluid. The radius of gyration $r_g$ was defined according to (Ellington 1984) (2.1). The two $Re$ are comparable to those of fruit flies (*Drosophila melanogaster*) (Dickinson *et al.* 1999; Birch *et al.* 2004; Lentink & Dickinson 2009b; Harbig *et al.* 2013b) and house flies (*Musca domestica*), respectively (Lentink & Dickinson 2009b; Harbig *et al.* 2013b). The kinematic viscosity used in the simulations was 8.0 cSt.

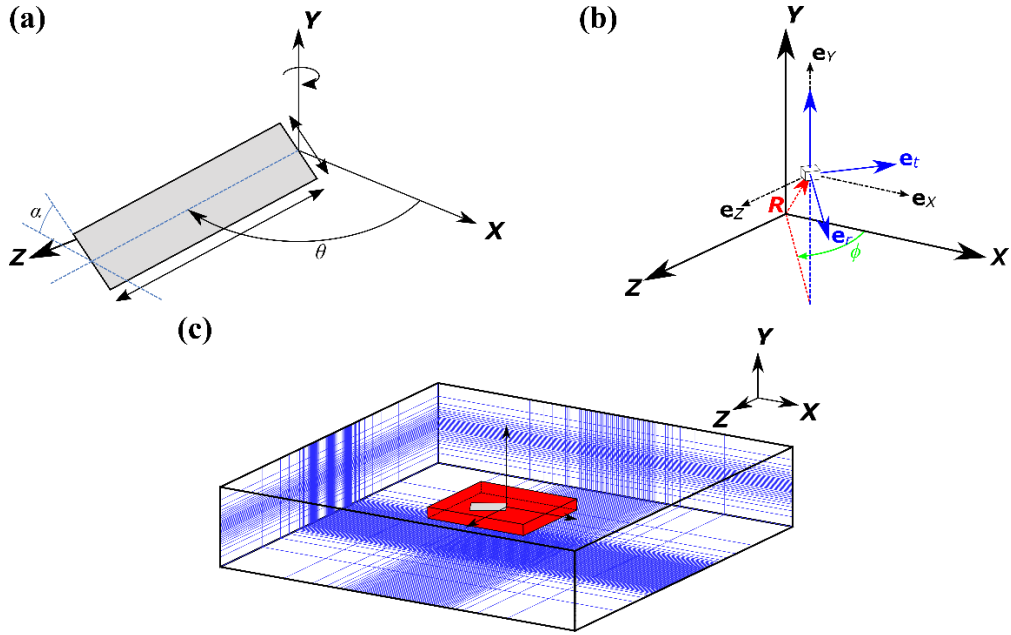

**Figure 2**. a) Wing kinematics and geometry. b) Definition of the rotated Cartesian frame $(t, y, r)$ using the azimuthal angle $\phi$ of a fluid particle measured from the fixed Cartesian frame $(X, Y, Z)$. c) Relative size of the grid chosen (blue) with the denser region (red) compared to the size of the wing. The rectangular domain has the size 50c x 10c x 50c with a dense mesh region in the center surrounded by the stretched meshes.

$$r_g = \left(\frac{\int_0^s c(r) r^2 dr}{\int_0^s c(r) dr}\right)^{\frac{1}{2}} = \frac{s}{\sqrt{3}} \qquad (2.1)$$

For the purpose of data analysis, in addition to a fixed Cartesian frame (X, Y, Z) a rotated Cartesian frame (t, y, r) was defined based on the azimuthal angle $\phi$ of a fluid element (see figure 2b), identical to those used in Cheng *et al.* (2013). Vectors in the fixed Cartesian frame were transformed into the rotated Cartesian frame using the Jacobian matrix $J(\phi)$.

$$J(\phi) = \begin{pmatrix} \sin\phi & 0 & -\cos\phi \\ 0 & 1 & 0 \\ \cos\phi & 0 & \sin\phi \end{pmatrix} \qquad (2.2)$$

For example, the fluid velocity and velocity gradients are transformed according to

$$\mathbf{u}(\mathbf{R}(t, y, r)) = J(\phi)\mathbf{u}(\mathbf{R}(x, y, z)), \qquad (2.3a)$$

$$\nabla_{t,y,r}\,\mathbf{u}(\mathbf{R}(t, y, r)) = J(\phi)\nabla_{x,y,z}\,\mathbf{u}(\mathbf{R}(x, y, z))J(\phi)^{-1} \qquad (2.3b)$$

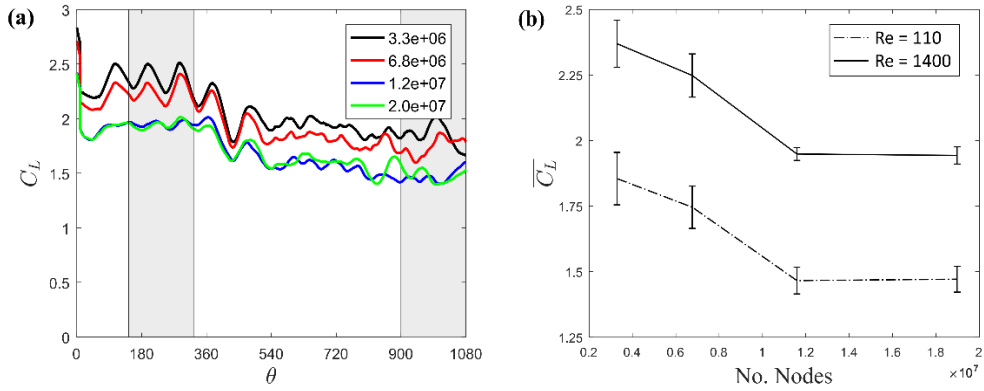

**Figure 3**. Four mesh sizes are used with 3.3, 6.8, 11.6, and 20 million nodes respectively. The black vertical lines correspond to the two time-veraging periods, $\theta = 144 - 324°$ and $900 - 1080°$ respectively. a) $C_L$ plotted against revolution angle $\theta$ for $AR = 5$, $Re = 1400$ simulation for each of the four mesh sizes tested. b) Time-averaged lift coefficient $\overline{C_L}$ plotted against numbers of nodes in each mesh with error bars representing the standard deviation $\sigma$ of the lift coefficient in each time-averaging period. Notice during both time-averaging periods that the $\overline{C_L}$ has converged to a single value between 11.6 and 20 million nodes.

where **R** is the position vector of a fluid element measured from the intersection of the origin and the axis of rotation.

A Cartesian computational grid with a stretching grid configuration was employed in the simulations, as shown in figure 2c. A grid spacing of 0.034c was used in the dense region for each of the three $AR$ values, which was sufficient to resolve the near-field vortex structures around the wing. A homogeneous Neumann boundary condition in pressure is applied to all six boundaries of the computational domain so that the vorticity could advect freely at the boundaries and a no-slip boundary condition was applied at the wing surface.

The governing equations employed by the solver are the N-S equations and incompressibility condition:

$$\frac{\partial \mathbf{u}}{\partial \tau} = -(\mathbf{u} \cdot \nabla)\mathbf{u} - \frac{1}{\rho}\nabla p + \nu \nabla^2 \mathbf{u}, \quad (2.4a)$$

$$\nabla \cdot \mathbf{u} = 0. \quad (2.4b)$$

Here, **u** is the velocity vector in the inertial reference frame (not from a perspective moving at the velocity of the wing), $\tau$ represents time, $\rho$ is fluid density, and p is pressure. Gravity was not included in the N-S equations solved here. An in-house, finite-difference-based, Cartesian-grid, immersed-boundary-method solver (Mittal *et al.* 2008) was employed to solve the above equations. In this solver, the flow simulation with complex moving boundaries was achieved with stationary non-body-conformal Cartesian grids to eliminate the need for a complex re-meshing algorithm, which was otherwise used by body-conformal methods. The solver can simulate flows of moving bodies with intricate geometry while still achieving second-order accuracy in both space and time. The equations were integrated in time using the fractional step method, and the boundary conditions on the immersed boundary were enforced by a ghost-cell procedure. This approach was successfully applied to the flapping propulsion of insects (Liu *et al.* 2016; Li & Dong 2017), birds (Ren *et al.* 2016), and fish (Liu *et al.* 2015, 2017). A detailed description and validation of this solver can be found in the authors' previous work (Dong *et al.* 2006; Li & Dong 2017).

A convergence test was performed using the lift coefficient data for four separate mesh sizes at an $AR = 5$, and $Re = 1400$. These four mesh sizes were 3.3 million nodes (225 x 65 x 255), 6.8 million (289 x 81 x 289), 11.6 million (353 x 93 x 353), and 20 million (337 x 113 x 337), respectively. Notice that the lift coefficient ($C_L$) was approximately the same between 11.6 and 20 million nodes (see figure 3a). There was negligible difference between the 11.6 and 20 mesh in the first and fourth revolution (see figure 3b). This indicates that the simulation had converged and that an 11.6 million node mesh size could be used at lower computational cost.

The aerodynamic forces acting on the wings are computed by the direct integration of the pressure and shear on the surfaces (Mittal *et al.* 2008). The lift force in particular is normalized as the lift coefficient $C_L = F_L / \frac{1}{2}\rho(\Omega r_g)^2 sc$ where, $F_L$ is lift force along the vertical direction (Y-positive).

## 2.2 *Analysis of Vorticity Dynamics*

To reveal the effects of the PVTr (i.e., planetary vortex tilting or the curl of Coriolis acceleration in the radial direction) on the vorticity dynamics, the flow data were cast into a relative rotating reference frame and used to evaluate the components of the relative vorticity equation. The relative vorticity equation (2.5) was arrived at by taking the curl of equation 2.4a and casting it into the relative rotating frame. The details of the derivation are provided in Kundu *et al.* (2008) and Cheng *et al.* (2013).

$$\frac{\partial \boldsymbol{\omega}'}{\partial \tau} = -(\mathbf{u}' \cdot \nabla)\boldsymbol{\omega}' + (\boldsymbol{\omega}' \cdot \nabla)\mathbf{u}' + (2\boldsymbol{\Omega} \cdot \nabla)\mathbf{u}' + \nu\nabla^2 \boldsymbol{\omega}' \qquad (2.5)$$

Here $\frac{\partial \boldsymbol{\omega}'}{\partial \tau}$ represents the rate of change of relative vorticity, $(\mathbf{u}' \cdot \nabla)\boldsymbol{\omega}'$ is the advection of relative vorticity, $(\boldsymbol{\omega}' \cdot \nabla)\mathbf{u}'$ is the tilting and stretching of relative vorticity, and $(2\boldsymbol{\Omega} \cdot \nabla)\mathbf{u}'$ is the tilting of planetary vorticity, and $\nu\nabla^2\boldsymbol{\omega}'$ is the molecular diffusion of relative vorticity.

Note that there is no planetary vortex tilting term $(2\boldsymbol{\Omega} \cdot \nabla)\mathbf{u}'$ in the vorticity equation in the inertial frame (Kundu *et al.* 2008; Cheng *et al.* 2013). This tilting term is identical to the curl of the Coriolis acceleration from the relative N-S equations, that is $(2\boldsymbol{\Omega} \cdot \nabla)\mathbf{u}' = -\nabla \times (2\boldsymbol{\Omega} \times \mathbf{u}')$. In a relative rotating frame, the planetary vortex is equivalent to twice the angular velocity of the rotating body, that is $\nabla \times (\boldsymbol{\Omega} \times \mathbf{R}) = 2\boldsymbol{\Omega}$ and the relative vorticity $\boldsymbol{\omega}' = \boldsymbol{\omega} - 2\boldsymbol{\Omega}$ ($\boldsymbol{\omega}$ being the vorticity in the inertial vorticity equation) (Kundu *et al.* 2008; Pedlosky 2013). For the current analysis, the planetary vortex was everywhere aligned with the axis of rotation of the wing (figure 1). Note that, in addition to the PVTr, there also exists the tilting of relative vorticity $\boldsymbol{\omega}'$.

Since the LEV vorticity is aligned with the radial direction, the analysis below focused exclusively on the radial component of the relative vorticity equation.

$$\frac{\partial \omega'_r}{\partial \tau} = -(\mathbf{u}' \cdot \nabla)\omega'_r + (\boldsymbol{\omega}' \cdot \nabla)u'_r - 2\Omega \frac{\partial u'_r}{\partial y} + \nu \nabla^2 \omega'_r \qquad (2.6)$$

The time derivative of radial vorticity (W) was evaluated indirectly using the sum of the terms on the RHS of 2.6. The spatial derivatives were calculated using a second-order, central-differencing scheme.

The first term on the RHS of 2.6 is the radial vorticity advection ($A$), which represents the transport of radial vorticity caused by the velocity field $\mathbf{u}'$. It can be decomposed into three contributions representing the advection along tangential, vertical, and radial directions:

$$A = -(\mathbf{u}' \cdot \nabla)\omega'_r = -\left(u'_t \frac{\partial}{\partial t} + u'_y \frac{\partial}{\partial y} + u'_r \frac{\partial}{\partial r}\right)\omega'_r. \qquad (2.7)$$

The second term (2.8a) represents the combined effect of tilting ($T$) and stretching ($S$) of relative vorticity in the relative rotating frame. This is decomposed into the tilting (2.8b) and stretching of the relative vorticity (2.8c):

$$(\boldsymbol{\omega}' \cdot \nabla)u'_r = \left(\omega'_t \frac{\partial}{\partial t} + \omega'_y \frac{\partial}{\partial y} + \omega'_r \frac{\partial}{\partial r}\right)u'_r, \qquad (2.8a)$$

$$T = \left(\omega'_t \frac{\partial}{\partial t} + \omega'_y \frac{\partial}{\partial y}\right)u'_r, \qquad (2.8b)$$

$$S = \omega'_r \frac{\partial u'_r}{\partial r}. \qquad (2.8c)$$

The tilting represents the vorticity reoriented from the other two directions into the radial direction by the gradients of radial flow in either the tangential or vertical direction. The stretching corresponds to the elongation of radially-oriented vortex tubes by the radial gradients of radial flow.

The third term on the RHS equation of 2.6 is the radial planetary vortex tilting or PVTr, represented here as $P$, which is simply the radial component of the planetary vortex tilting/stretching described above. Note that the PVTr scales with the vertical gradient of radial velocity (2.9):

$$P = -2\Omega \frac{\partial u'_r}{\partial y}. \qquad (2.9)$$

The last term on the RHS of equation 2.6 is the diffusion or dissipation of radial vorticity ($D$), which represents the transport of vorticity out of the LEV due to the molecular interactions (Batchelor 2000):

$$D = \nu \nabla^2 \omega'_r = \nu \left( \frac{\partial^2}{\partial t^2} + \frac{\partial^2}{\partial y^2} + \frac{\partial^2}{\partial r^2} \right) \omega'_r. \quad (2.10)$$

Previous experimental results (Cheng *et al.* 2013) showed that the negative radial vorticity comprising the LEV was continually generated at the leading-edge and advected tangentially into the wake region as the wing rotated. The vorticity was further advected downward into the wake by vertical flow, where there exists considerable stretching and tilting that reduces LEV vorticity. The vorticity transport due to the spanwise flow, however, was found to be negligible.

A normalization scheme was then applied using the following characteristic values: the velocity by the speed of the wing at its radius of gyration $\mathbf{u}^* = \frac{\mathbf{u}}{\Omega r_g}$, the vorticity by the magnitude of the planetary vortex $\boldsymbol{\omega}^* = \frac{\boldsymbol{\omega}}{2\Omega}$, and the length by the wing chord length $\nabla^* = c\nabla$. Rather than using an independent vorticity scale, Lentink & Dickinison (2009a) derived theirs from a reference velocity $\mathbf{u}^+ = \frac{\mathbf{u}}{\Omega s}$ and length $\nabla^+ = c\nabla$ giving $\boldsymbol{\omega}^+ = \frac{\boldsymbol{\omega}}{\Omega AR}$. This normalization scheme was used as an alternative to the first normalization scheme by Lentink & Dickinison (2009a). Normalized quantities using planetary vorticity $2\Omega$ were denoted by a superscript $^*$, while those using $\Omega AR$ were denoted by a superscript $^+$.

The $'$ used for relative quantities was dropped from here onwards. Both sets of normalization schemes arrived at the same normalized PVTr, $P^*$ (2.11a) but different normalized advection, $A^*$ (2.11b) and $A^+$ (2.11c) as functions of the $Ro$ for fixed $Re$ (defined in §2.1). These relationships were arrived at by noting that with a constant $Re$, the angular velocity $\Omega$ scales inversely with the $Ro$.

$$P = P^* \left( \frac{2\nu^2 Re^2}{c^4 Ro} \right) \propto P^* \cdot Ro^{-1} \quad (2.11a)$$

$$A = A^* \left( \frac{2\nu^2 Re^2}{c^4 Ro} \right) \propto A^* \cdot Ro^{-1} \quad (2.11b)$$

$$A = A^+ \left( \frac{\nu^2 s^2 Re^2}{c^4 r_g^2} \right) \propto A^+ \cdot Ro^0 \quad (2.11c)$$

The $Ro$ is defined as the ratio of the advective and Coriolis accelerations (2.12a) which is equivalent to the normalized radius of gyration (Kundu 2008; Pedlosky 2013). Additionally, a local Rossby number $Ro(r)$ (2.12b) was defined using the local velocity at a radial position r along the wing-span as described by Jardin & David (2017):

$$Ro = \frac{|(\mathbf{u}' \cdot \nabla)\mathbf{u}'|}{|2\mathbf{\Omega} \times \mathbf{u}'|} \propto \frac{r_g}{c}, \quad (2.12a)$$

$$Ro(r) = \frac{r}{c}. \quad (2.12b)$$

Assuming that $P^*$, $A^*$, and $A^+$ scale with $\mathcal{O}(1)$, the above normalization schemes indicate that with a constant $Re$, that the dimensional PVTr (P) is inversely proportional to $Ro$ (2.11a), while the dimensional advection (A) is inversely proportional to $Ro$ (2.11b) or independent of $Ro$ (2.11c) depending on the vorticity scale chosen. The relative effect of the PVTr can be further revealed by calculating the ratio of the advection with PVTr. Using the two normalization schemes, we have:

$$\frac{A}{P} = \frac{A^*}{P^*} \propto \frac{A^*}{P^*} Ro^0, \quad (2.13a)$$

$$\frac{A}{P} = \frac{A^+}{P^*} \frac{s^2 Ro}{2r_g^2} \propto \frac{A^+}{P^*} Ro^1. \quad (2.13b)$$

Under the assumption that both the normalized ratios scale with $\mathcal{O}(1)$, it is clear that the ratio A/P scales with $Ro^0$ (2.13a), or with $Ro^1$ (2.13b). Therefore, we make the prediction that the ratio A/P will scale with $Ro^n$, where $n = 0$ or 1. To test this prediction and to understand the dependency of PVTr on $Ro$ the value of n will be estimated based on the data from the numerical simulation for both $Re$ (see §3.3):

$$\frac{A}{P} \propto Ro^n. \quad (2.14)$$

## 2.3 *Averaging Processes*

To investigate the vorticity dynamics of the LEV during the quasi-steady period, two averages of the terms in equation 2.6 were taken. These are the time-average (denoted by $\overline{\phantom{a}}$ ) and the spatial-averages (denoted by $\widehat{\phantom{a}}$ , and $\widetilde{\phantom{a}}$ for the area, and volume-averages respectively). The time-average was calculated first, and then used to calculate the spatial-averages. Both averages were calculated within LEV control-volumes over the second half of the third revolution because the flow had reached an approximately quasi-steady state (see §3.1). The LEV control-volumes were identified using a threshold of the normalized time-averaged radial vorticity $\overline{\omega_r^*} = -3$, which was chosen instead of the more common *Q*-criterion (Jeong & Hussain 1995; Harbig *et al.* 2013a,b; Cheng *et al.* 2013; Garmann & Visbal 2014; Jardin & David 2015) or vorticity magnitude (Mao & Jianghao 2004; Cheng *et al.* 2013; Bos *et al.* 2013) because only radial vorticity dynamics were being considered. Other values were tested but they did not have any measurable effect on the results.

The process for calculating the time-average essentially collapsed all of the LEV control-volumes from each time-step onto a single wing location (in this case the original position of the wing) (figure S1a), and corrected the misalignment of the resulting grids (figure S1c). The correction process identified all the vorticity equation data on the misaligned grid (rotated by an angle $\phi$) within a small radius $\delta = 0.8\sqrt{(\Delta x)^2 + (\Delta z)^2}$ centered at each point on the original grid and averaged them together returning this new adjusted value to the original grid. From this point, all of this corrected data was averaged together to produce a time-averaged LEV control-volume (figure S1b,d).

Once the time-average was calculated, it was used to determine the spatial-averages for each $Re$ and $AR$. First, the area-averages were calculated along surfaces defined by the intersections of a set of cylindrical slices with radii equal to the $Ro(r)$ centered and aligned with the rotation axis and the interior of the time-averaged LEV (figure S1e). The number of cylindrical slices used were 21, 41, and 31 for $AR = 3$, 5 and 7 respectively. All of the vorticity equation data within a small tolerance of each slice were averaged together to calculate the area-average at each $Ro(r)$. The volume-averages were then calculated as the average of the area-averages.

Due to the infinitesimal thickness of the wing, most of the vorticity is produced at or very near the leading-edge of the wing. Consequently, large gradients in the velocity field exist in the regions close to the leading-edge. In these regions, finite differencing is subject to higher truncation errors because of the limited resolution of the grid or mesh in the CFD simulation, and this could lead to numerical diffusion of vorticity. This "false" numerical diffusion effect occurs because the local streamlines of the flow are not aligned with the local computational mesh. While numerical diffusion can be reduced by higher-order, finite-difference schemes and/or by more finely resolved meshes, the alignment issue remains since streamline geometry is not known in advance when generating the mesh. For a more detailed discussion about numerical diffusion, see (Patankar 1980). The effects of these errors can be seen by calculating the sum of the terms on the RHS of equation $\widehat{W}^*$ 2.6, which was expected to be close to zero after time and spatial averaging because the flow had reached quasi-steady state within the time-averaging period. However, this was not the case as is observed in figures S3a and S4a. The time and area-average of the time derivative of vorticity is not zero initially after taking the time-average. This "false" numerical diffusion can lead to errors in calculating the terms in equation 2.6 and can be reduced by using higher-order, finite-difference schemes and/or by more finely resolved meshes near the leading-edge. These methods are difficult to implement due to the nature of in-house CFD code used. Instead, a leading-edge cylinder (LEC blue) and tip cylinder (TC red) was used to isolate certain problematic points in the mesh and remove them from the time-averaged LEV control-volume (figure S2). No cylinders were used for the root or trailing edges since the LEV control-volume does not cover those parts of the wing. The result of using a LEC and TC is shown in figures S3b-d, and S4b-d. After applying the LEC, the time derivative became roughly zero everywhere as expected, although there are higher deviations when $Re = 1400$.

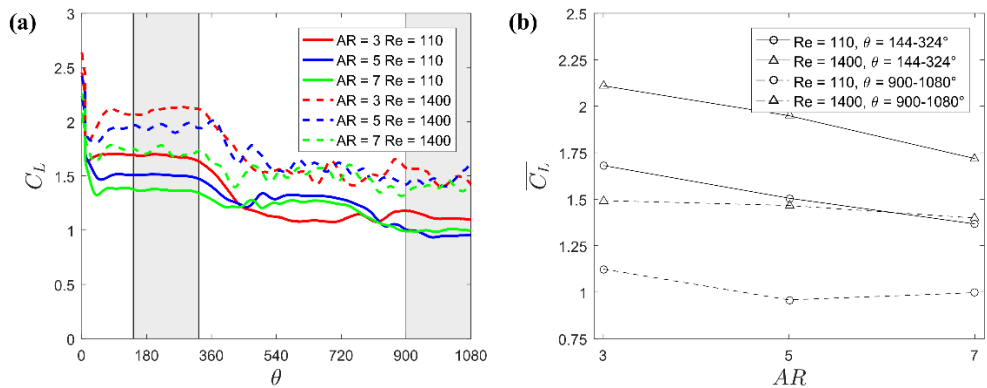

**Figure 4**. a) $C_L$ plotted against revolution angle $\theta$ for each case. b) $\overline{C_L}$ plotted against $AR$ for each $Re$. Solid lines represent the second half of the first revolution ($\theta = 144 - 324°$) and dashed lines represent the second half of the third revolution ($\theta = 900 - 1080°$). Triangles represent $Re = 110$, and circles represent $Re = 1400$.

## 3. Results and Discussion

### 3.1 *Vortex Structure and Lift Coefficient*

In this work, the vorticity dynamics of revolving wings were analysed only when the flow field (and therefore the LEV) had reached an approximately quasi-steady state in the relative rotating frame. The quasi-steady state was identified by first observing the temporal behaviours of $C_L$ and the vortex structure over the first three revolutions. Figure 4a shows these temporal variations of $C_L$ for all six cases with different $AR$ and $Re$ and figure 4b shows the time-averaged values in the second half of the first and third revolution, respectively. In all six cases, there was a transient period for approximately the first 90° of rotation before $C_L$ plateaued in the rest of the first revolution, after which all the curves had a similar stepwise decreasing behaviour. Within the first and third revolutions, the time-averaged $C_L$ increased with increasing $Re$, and decreased with increasing $AR$ (figure 4b). Based on the trend of $C_L$ (figure 4a) it is determined that the flow field has reached a quasi-steady state by the second half of the third revolution, which was the main time-period of our analysis.

Figure 5 shows iso-surfaces of normalized, and time-averaged radial vorticity. When $Re = 110$, the vortex structures were smoother and subject to smaller temporal variations when compared with the cases where $Re = 1400$. The LEV was comprised of negative radial vorticity and is conical in shape over the wing surface, extending past the trailing edge for the $AR = 5$, and 7 cases. The LEV sheds into a TiV near the tip and extended into the wake aligned with the positive tangential direction.

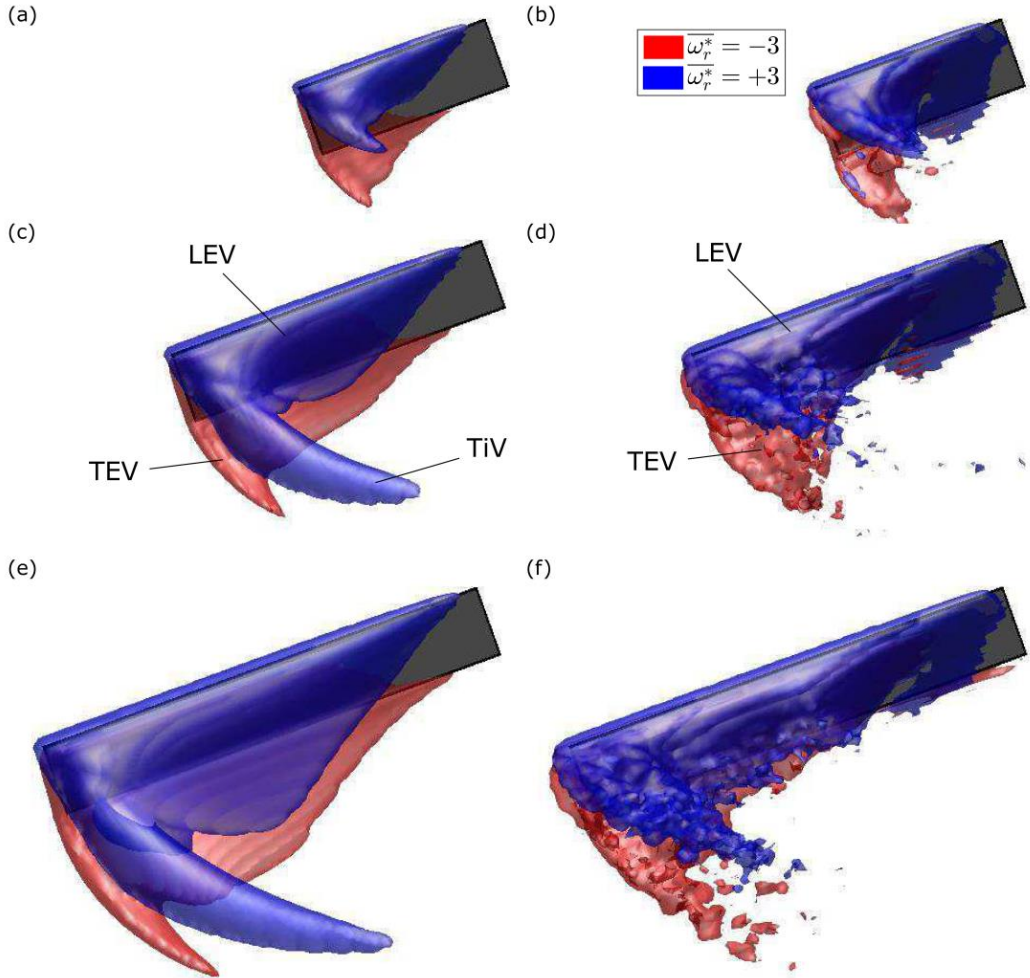

**Figure 5**. Normalized time-averaged radial vorticity iso-surfaces: a) $AR = 3, \ Re = 110$, b) $AR = 3, \ Re = 1400$, c) $AR = 5, \ Re = 110$, d) $AR = 5, \ Re = 1400$, e) $AR = 7, \ Re = 110$, f) $AR = 7, \ Re = 1400$.

## 3.2 *Local Behaviour and Effects of the PVTr*

The PVTr represents the effect of the Coriolis acceleration in the vorticity dynamics (2.6) as discussed in §2.2. The PVTr is directly proportional to the vertical gradient of the radial velocity (2.9). Earlier we predicted that the PVTr acted functionally as a source of radial vorticity opposite to that of the LEV by tilting the planetary vortex line into the radial direction. This mechanism was examined using the area-averages of PVTr within the LEV control-volume for all six cases.

Figure 6 shows the area-averaged values of the PVTr $P$ (2.9) along with the advection $A$ (2.7), tilting $T$ (2.8b), stretching $S$ (2.8c), and diffusion $D$ (2.10) as they vary with $Ro(r)$. It was observed that the PVTr was consistently positive for the entire length of the wing. Since the PVTr is always positive, it is generating oppositely-signed radial LEV vorticity, and therefore acts to reduce the LEV circulation for all $Re$ and $AR$ tested. For the $Re = 110$ cases, the strongest positive term over the majority of the wing-span was the PVTr. In contrast, when $Re = 1400$, the PVTr was the second strongest positive term following the vortex tilting except when $AR = 7$ where the PVTr and tilting had similar magnitudes. Therefore, the stabilizing effect of the PVTr at $Re = 1400$ is weaker relative to it's effect at $Re = 110$ because the tilting plays an additional role in LEV stabilization (see §3.5). Additionally, the PVTr exhibited a roughly linear behaviour up until approximately 1-2 chord lengths from the tip where it reached the spanwise location of maximum PVTr for all $Re$ and $AR$ tested, which was identified as $Ro(r_N) = \arg\max(\widehat{P^*})$. The values of $\max(\widehat{P^*})$ and $AR - Ro(r_N)$ are provided in table 1. For all $AR$ and $Re$ tested, the maximum PVTr varied between approximately 1.5 and 2.5. This indicates that the magnitude of the PVTr does not change substantially with the changes in $AR$ or $Re$ and functions as a consistent mechanism contributing to LEV stability.

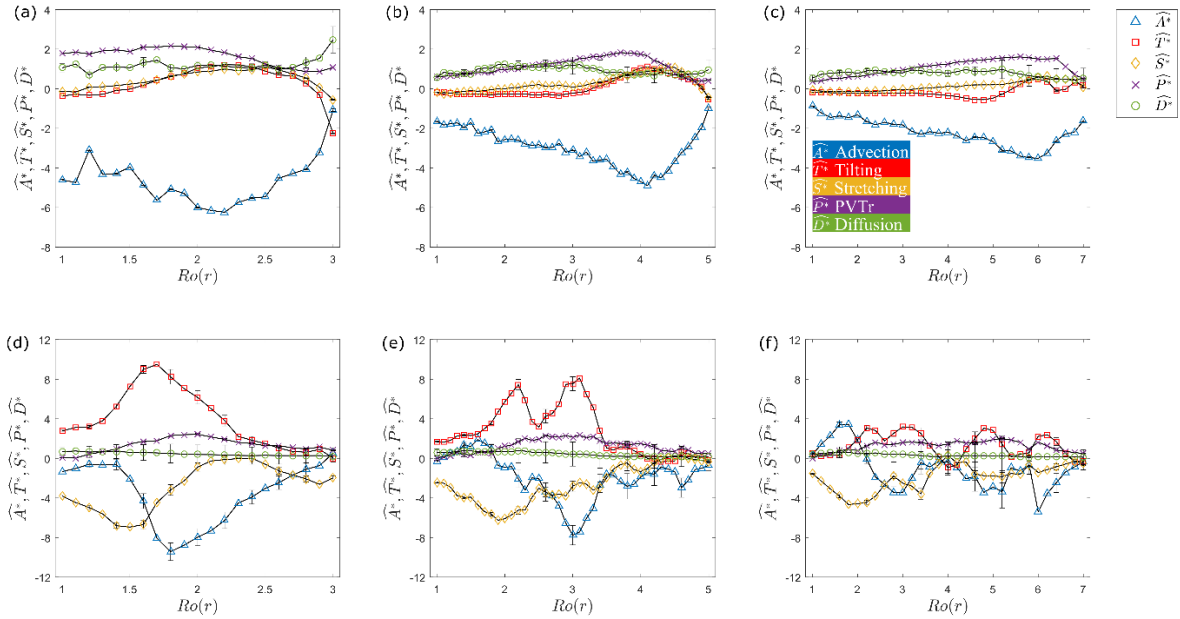

**Figure 6.** Time and spatially averaged terms of the non-dimensional radial vorticity equation (2.6). The different terms shown are the advection $\widehat{A^*}$ (2.7) blue, tilting $\widehat{T^*}$ (2.8b) red, stretching $\widehat{S^*}$ (2.8c) yellow, radial planetary vortex tilting (PVTr) $\widehat{P^*}$ (2.9) purple, and diffusion $\widehat{D^*}$ (2.10) green. Error bars representing one standard deviation from the mean are also provided. a) $AR = 3, Re = 110$, b) $AR = 5, Re = 110$, c) $AR = 7, Re = 110$, d) $AR = 3, Re = 1400$, e) $AR = 5, Re = 1400$. Note that the PVTr is consistently positive across all $AR$ and $Re$ for the majority of the wing-span. When $Re = 110$ the PVTr and the advection both scale with the $Ro(r)$ up to one chord length from the tip. When $Re = 1400$, the tilting now exceeds the PVTr and scales roughly with the advection.

| AR | $Re = 110$ | | $Re = 1400$ | |
| --- | --- | --- | --- | --- |
| | $\max(\widehat{P^*})$ | $AR - Ro(r_N)$ | $\max(\widehat{P^*})$ | $AR - Ro(r_N)$ |
| 3 | 2.163 | 1.2 | 2.2428 | 1.0 |
| 5 | 1.830 | 1.3 | 2.338 | 1.9 |
| 7 | 1.603 | 0.4 | 2.006 | 1.8 |

**Table 1.** Maximum $\widehat{P^*}$, and length of non-linear region of $\widehat{P^*}$

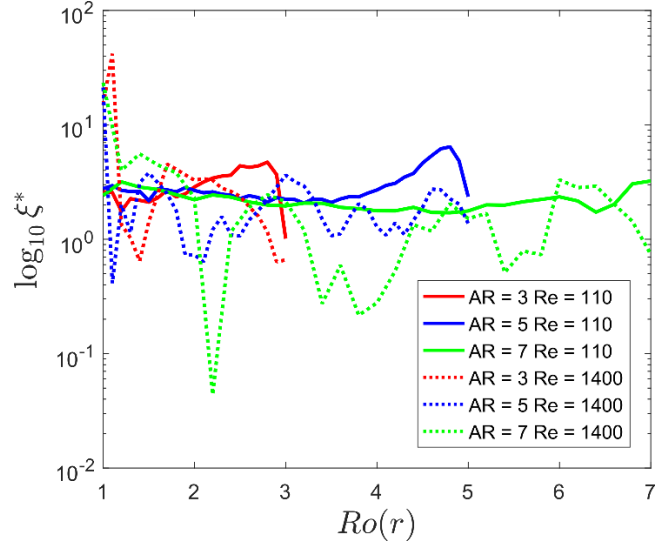

**Figure 7**. Local ratio of advection to PVTr ($\xi^*$) plotted against $Ro(r)$ on a semiology-y plot. When the $Re = 110$, this ration remains approximately $\mathcal{O}(1)$. When $Re = 1400$, the ratio exhibits large fluctuations along the wing-span.

The primary term from the vorticity equation (2.6) that leads to the growth and instability of the LEV was the vorticity advection (2.7). As the wing rotated, the tangential flow that passed over the leading-edge transports negative radial vorticity generated at the fluid-surface interface into the LEV. Therefore to further understand the relative stabilizing effect of the PVTr, the ratio of the advection to the PVTr using area-averages $\xi^* = \left|\widehat{A^*}/\widehat{P^*}\right|$ was calculated (see 2.11b and §2.3). A plot of $\xi^*$ vs. $Ro(r)$ is given in figure 7 for each $AR$ and $Re$ tested. In general, $\xi^*$ remained approximately constant at values of $\mathcal{O}(1)$ across the wing-span when $Re = 110$, indicating that the relative effect of PVTr in the vorticity dynamics remains consistent along the wing-span (except the region close to the tip). In the cases of $Re = 1400$ however, $\xi^*$ exhibited large variations along the wing-span and the effect of PVTr was less dominant both locally and globally. Presumably, the onset of turbulence in the $Re = 1400$ leads to large spatial variations in the flow velocity, which significantly increases the variability in the vorticity advection (figures 6d-f).

### 3.3 *Global Behaviour and Effects of AR and Re on the PVTr and Vorticity Advection*

The effect of the PVTr on the entire LEV control-volume was determined by considering the normalized volume-averages of PVTr for each case (figure 8a). In this section, the volume-averaged PVTr $\widetilde{P^*}$ was first plotted against different $AR$ for $Re = 110$ and 1400 (figure 8a). With the volume-averaged advection $\widetilde{A^*}$, the ratio between the advection and PVTr $\Xi^* = \left|\widetilde{A^*}/\widetilde{P^*}\right|$

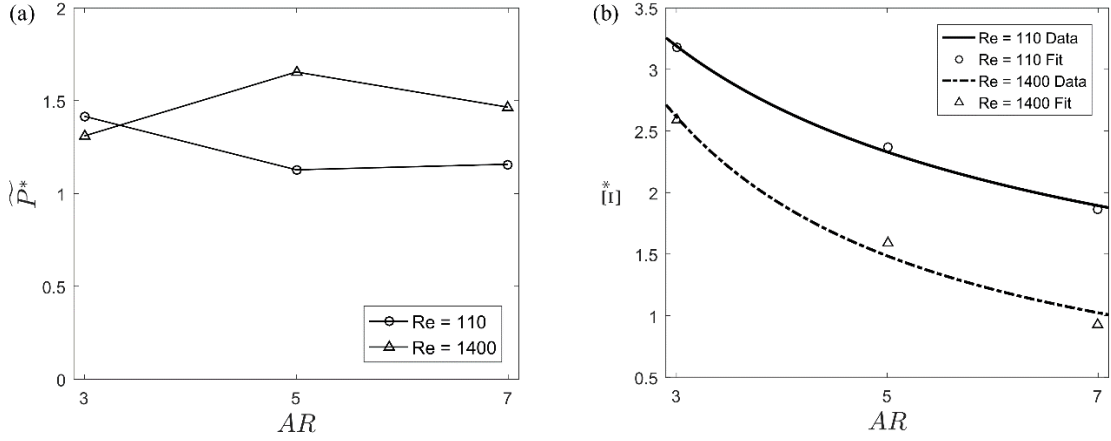

**Figure 8**. a) volume-averages of the PVTr plotted against $AR$ for $Re$ = 110 and 1400. The volume-averaged PVTr is observed to remain approximately constant across all AR and Re at values. b) Global ratio of advection to PVTr $\Xi^*$ plotted against $AR$ for $Re$ = 110 and 1400. A factor of $(s/r_g)^n$ is used to scale back $\Xi^*$ to vary with the $Ro$ as in equation $\Xi^*$. The global ratio exhibits a hyperbolic decrease as shown by the curve fits, and decreases with increasing $Re$.

(figure 8b) was then calculated, similar to how $\xi^*$ was calculated for the local PVTr $\widehat{P^*}$ and advection $\widehat{A^*}$ (see §3.2).

The volume-averaged PVTr remained approximately $\mathcal{O}(1)$ for all of the $AR$ and $Re$ tested. This was similar to what was observed regarding the magnitude of the area-averaged PVTr $\widehat{P^*}$ in §3.2. This further evidences that the PVTr consistently generates oppositely-signed vorticity within the entire LEV control-volume and therefore plays a key role in the stability of the LEV.

The stabilizing effect of the PVTr in the entire LEV was further quantified using the ratio of the volume-averages of advection and PVTr, $\Xi^* = |\widetilde{A^*}/\widetilde{P^*}|$ (see §2.3). The volume-averaged ratio $\Xi^*$ showed a relatively consistent behaviour with the $AR$ across both $Re$ values. That is, $\Xi^*$ decreased with increasing $AR$ i.e. n < 0 where n is the exponent in equation 2.14 (see figure 8b). Furthermore, $\Xi^*$ also dropped in magnitude with increasing $Re$. Due to the hyperbolic dependency on $AR$ (or $Ro$ since $AR$ scales with $Ro$) predicted by the normalization analysis, the dependency of $\Xi^*$ with $AR$ was modeled using a hyperbolic fit for both values of $Re$ with the $AR$ as the independent variable (3.1).

$$\Xi^*(Ro) = aRo^n \tag{3.1}$$

The values for a, and n as well as a measure of the accuracy of the fit are provided in table 2. Since n < 0, neither of the two normalization schemes, which predicted n = 0 or 1, successfully predict the dependency of $\Xi^*$ with $Ro$. This failure could result from the scaling of either advection or PVTr being inappropriate. However, since the magnitude of the non-dimensional PVTr $\widetilde{P^*}$ is of $\mathcal{O}(1)$ (see figure 8a), the failure must lie with the non-dimensional advection $\widetilde{A^*}$. Due to the small parameter space of $AR$ and $Re$ involved in this study, it is unclear what the cause of this failure in the advection might be. In future work, a widened parameter space, and decoupling the $Ro$ and $Re$ (Harbig *et al.* 2013b) might be able to help realize this fundamental relationship between the advection, PVTr, $Ro$, and $Re$.

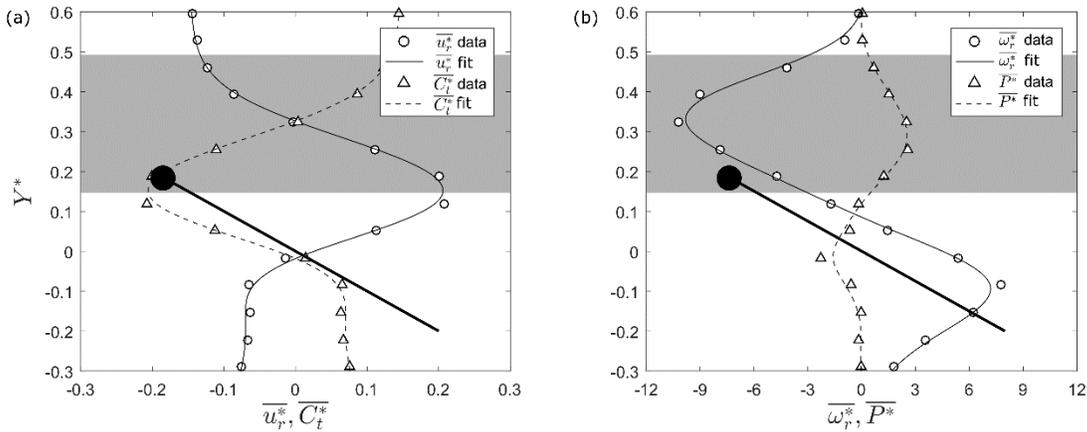

**Figure 9**. The LEV control-volume is defined by $\overline{\omega_r^*} \leq -3$ outlined in gray, and the wing cross-section is visualized by a black line. The wing's root is aligned with rotation axis at the mid-chord location. The circles and triangles represent the actual data, while the solid and dashed lines are simply spline fits of the data. A black circle is used to specify the wing leading-edge. a) Normalized radial velocity and tangential Coriolis acceleration plotted against normalized vertical position. Note the positive vertical gradient in the radial velocity and tangential Coriolis acceleration is directly aligned with the LEV control-volume. b) Normalized radial vorticity and PVTr plotted against normalized height above the wing. Note that the region of positive PVTr is directly aligned with the LEV control-volume.

|                    | $\Xi^*$    |             |
| ------------------ | ---------- | ----------- |
| Coeff./Qual. of fit | $Re = 110$ | $Re = 1400$ |
| $a$                | 4.472      | 4.802       |
| $n$                | -0.616     | -1.107      |
| $R^2$              | 0.998      |             |

**Table 2**. Coefficients for the hyperbolic curve fits based on the $Ro$.

### 3.4 *The Relationship of Coriolis acceleration, PVTr, and LEV Vorticity*

This section provides further evidence regarding the underlying physics of the PVTr by examining the relationships among radial flow, Coriolis acceleration, PVTr, and radial (LEV) vorticity. This was accomplished by analyzing the flow data along vertical lines at the middle of the wing-span ($Ro(r) = 2.5$) for the $AR = 5$ and $Re = 110$ case. Here it is shown that the peak in the PVTr directly coincides with the peak of the LEV vorticity.

The Coriolis acceleration, non-dimensionalized by the magnitude of the planetary vortex and the velocity at the radius of gyration, is given by equation 3.2:

$$\boldsymbol{C} = \begin{pmatrix} -2\Omega u'_r \\ 0 \\ 2\Omega u'_t \end{pmatrix} = \begin{pmatrix} C_t \\ 0 \\ C_r \end{pmatrix}. \tag{3.2}$$

Figure 9a shows the time-averaged non-dimensional tangential Coriolis acceleration $\overline{C_t^*}$ and radial velocity, and the gray highlighted region represents the region of the LEV defined using the non-dimensional radial vorticity in figure 9b. There exists a negative gradient in the radial velocity in the region of the LEV, and therefore a positive gradient of tangential Coriolis acceleration along the vertical direction. The positive gradient of tangential Coriolis acceleration would create angular acceleration of the fluid particles (along the positive radial direction) opposite to the direction of LEV vorticity and therefore reduce its strength. Since the PVTr is identical to the vertical gradient of Coriolis acceleration (2.9), it produces oppositely-signed vorticity to the LEV (figure 9b). Figure 9b shows that the PVTr acts in the opposite direction of LEV vorticity, therefore contributing to limiting its growth.

### 3.5 *Local Vorticity Dynamics*

The local behaviours of other terms in the vorticity equation in addition to the PVTr are discussed here, according to their area-averages. For the $Re = 110$ cases (figure 6a-c), the advection (blue) was negative for the entire length of the wing, and increased linearly along the wing-span until approximately one chord length from the tip (or $Ro(r) = AR - 1$), although this was more apparent for the $AR = 5$ and 7 cases (figure 6b-c). The sign of the advection was expected, since the advection works as a primary source of LEV vorticity (negative radial vorticity). The linear behavior in the advection can be explained by considering that locally, the angular velocity is constant, and therefore the non-dimensional advection scales with $Ro(r)$. This comports with the decreasing slope of the advection curves with increasing $AR$ since for a fixed $Re$ the angular velocity also decreases. Finally, the abrupt change in the linear behavior of the advection close to the wing tip is likely due to the presence of a TiV. In the TiV, the radial vorticity is being tilted predominantly into the tangential direction. Unlike the advection, the tilting $\widehat{T^*}$ (red) and stretching $\widehat{S^*}$ (yellow) started out at approximately zero and remain approximately constant for the majority of the wing-span at $Re = 110$, where both increased slightly within the tip region before decreasing again (figure 6b-c). Since their values were approximately zero for the majority of the wing-span, it is likely that they do not play a significant role in the stability of the LEV outside of the tip region at this $Re$. Finally, for $Re = 110$, the diffusion $\widehat{D^*}$ (green) was approximately constant for the entire wing-span, but with a higher variability over time (with larger error bars). Additionally, the diffusion was the second largest positive term immediately following the PVTr, and therefore is not negligible in the vorticity dynamics. Note that, Jardin (2017) showed that the role of the diffusion in LEV stability becomes significant for $Re < 200$. Together these results suggest that when $Re = 110$, the diffusion and the PVTr are both stronger contributors to LEV stability.

At $Re = 1400$, the advection was still negative, and had significantly higher nonlinearity and variability (see figure 6d-f) especially at higher $AR$. This suggests that the linear behaviour observed in the advection is $Re$ dependent. A minimum exists at approximately the mid-span location for the $AR = 3$ and 5 cases, but for the $AR = 7$ case no such trend exists. The tilting roughly mirrored the advection, having one or more peaks coinciding with the minima of the advection for $AR = 3$, and 5 and significantly dropping in magnitude for $AR = 7$. This behaviour suggests that at higher $Re$ the tilting becomes a stronger contributor to LEV stability that balances the advection, especially at $AR = 3$, and 5 when the magnitude of tilting is significantly greater than that of the PVTr. The stretching was consistently negative and at its lowest point was of the same order of magnitude of the minimum advection. This suggests that at $Re = 1400$ it has a similar role as advection and tends to destabilize the LEV. Finally, the diffusion continues to be roughly constant along the entire wing-span, but was now much lower than all the other terms and therefore does not contribute significantly to the stability of the LEV at a $Re = 1400$. Together, these results suggest that although the PVTr remains at about the same magnitude across all $AR$ and $Re$, again confirming it is a consistent mechanism contributing to LEV stability, its relative contribution to LEV stability seems to be $Re$ and $AR$ dependent due to the different behaviors of advection, tilting and stretching and vorticity dynamics.

### 3.6 *Global Vorticity Dynamics*

Figure 10 and table 3 show the volume-average data for each of the terms on the RHS of equation 2.6. The values are presented as a bar graph in figure 10 to compare the six cases. When $Re = 110$, the advection had the strongest magnitude and decreased with increasing $AR$. Both the tilting and stretching were approximately negligible compared to the other terms at this $Re$, although the stretching was slightly larger than the tilting. The PVTr, as was discussed earlier, was the largest positive term and remained approximately constant across all $AR$ and $Re$. The diffusion was less than the PVTr, and approximately negligible for $AR = 3$, and 7 but is approximately half the size of the PVTr when $AR = 5$. When $Re$ increases to 1400, the advection behaved similarly to the $Re = 110$ cases, decreasing with increasing $AR$ but slightly in magnitude. The tilting, on the other hand, increased significantly and was approximately equal to the PVTr when $AR = 5$. Similarly, the stretching also increased in magnitude but is now negative for all $AR$. The magnitude of the stretching approaches a similar value as the advection when $AR = 5$. The effects of tilting and stretching on the LEV stability, as they become more dominant at $Re = 1400$, remain to be studied further in the future work. Finally, the diffusion became negligible when $Re = 1400$ just as the tilting and stretching were when $Re = 110$.

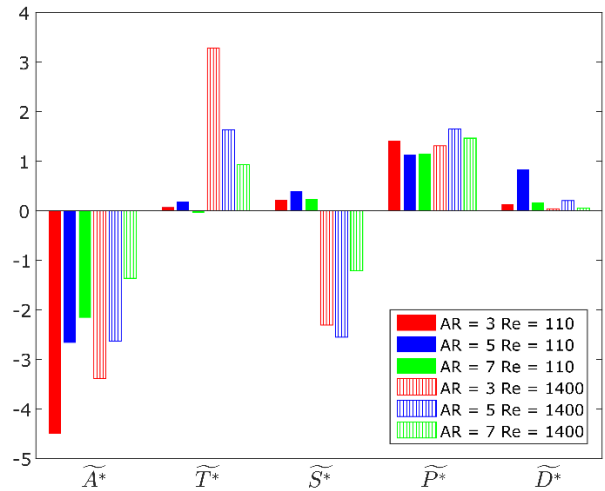

**Figure 10**. Volume-averaged vorticity dynamics. The advection is consistently the highest in magnitude. The tilting, stretching, and diffusion are for the most part negligible at $Re = 110$ except for $AR = 5$ where the diffusion exceeds both the tilting and the stretching. At $Re = 1400$, the tilting is approximately equal to the PVTr at $AR = 5$. The stretching changes sign when $Re = 1400$, and exceeds the magnitude of the PVTr at $AR = 5$. The diffusion remains positive but is mostly negligible at $Re = 1400$.

| AR | $Re = 110$ | | | | | $Re = 1400$ | | | | |
|---|---|---|---|---|---|---|---|---|---|---|
| | $\widetilde{A^*}$ | $\widetilde{T^*}$ | $\widetilde{S^*}$ | $\widetilde{P^*}$ | $\widetilde{D^*}$ | $\widetilde{A^*}$ | $\widetilde{T^*}$ | $\widetilde{S^*}$ | $\widetilde{P^*}$ | $\widetilde{D^*}$ |
| 3 | -4.500 | 0.071 | 0.214 | 1.416 | 0.134 | -3.393 | 3.283 | -2.308 | 1.310 | 0.042 |
| 5 | -2.668 | 0.187 | 0.393 | 1.128 | 0.836 | -2.633 | 1.602 | -2.554 | 1.654 | 0.204 |
| 7 | -2.160 | -0.040 | 0.238 | 1.157 | 0.171 | -1.363 | 0.933 | -1.213 | 1.465 | 0.058 |

**Table 3**. Volume-averaged data for each term in the vorticity equation.

## 4. Conclusions

This work investigated the local and global vorticity dynamics within the LEV by simulating the flow over six revolving wings with $AR = 3$, 5, and 7 and $Re$ at the radius of gyration of 110 and 1400, respectively, using an in-house, finite-difference-based, immersed-boundary-method solver. The data within the LEV control-volume were time-averaged over the second half of the third revolution after identifying that the flow had reached a quasi-steady state. We observed that the radial component of the curl of the Coriolis acceleration, or the PVTr, provided a relatively consistent stabilizing effect against the vorticity advection regardless of the value of $AR$ or $Re$. This PVTr varied linearly with $Ro(r)$, increasing along the wing-span for all tested values of $Re$ and $AR$. The relative strength of the PVTr was observed to be inversely proportional to both $AR$ and $Re$ as well. The PVTr was notably effective at $Re = 110$ where it was the strongest positive term for the entire wing-span seconded by the vorticity diffusion. However, when $Re = 1400$, the tilting became the dominant stabilizing term for $AR = 3$ and 5, as the PVTr still behaved similarly to the $Re = 110$ cases. All of this leads us to conclude that the PVTr is a consistent mechanism that limits the growth of the LEV and contributes to its stability, its effect is particularly strong at $Re = 110$ but less significant compared with other three-dimensional effects (vortex tilting) at $Re = 1400$. Since the PVTr is the gradient of the Coriolis acceleration, it is clear that the Coriolis acceleration plays a key role in the LEV stability which is in agreement with Lentink & Dickinson (2009a,b); Jardin & David (2014, 2015) and Jardin (2017). However, the observation that the vorticity advection grew weaker with increasing $AR$ was not previously observed. Finally, the results show that the effects of PVTr and possibly other three-dimensional effects depend on both $Re$ and $Ro$, and that $Ro$ alone is insufficient to quantify their effects on the LEV stability in revolving wings.

Further work must be done to quantify the effect of angle of attack on the vortex dynamics. It is possible that the varying the angle of attack, for fixed $Re$ and $AR$, could fundamentally change how the PVTr aligns with the region of the LEV. Additionally, an analysis must be done to decouple the effect of $Re$ and $Ro$ over a wider parameter space for the rotating wing. A similar analysis was done by Harbig *et al.* (2013b), but this can also be done by fixing the $Re$ at a single point along the wing-span for multiple $AR$. This could help to better understand why the advection decreased with increasing $AR$ as was observed in this study.

## Acknowledgments

We would like to thank Roberta J. Werner, Long Chen, and Robin Barrett for reviewing the manuscript along with several others (who choose to remain anonymous).

The role of planetary vortex tilting on the stability of the leading-edge vortex    23

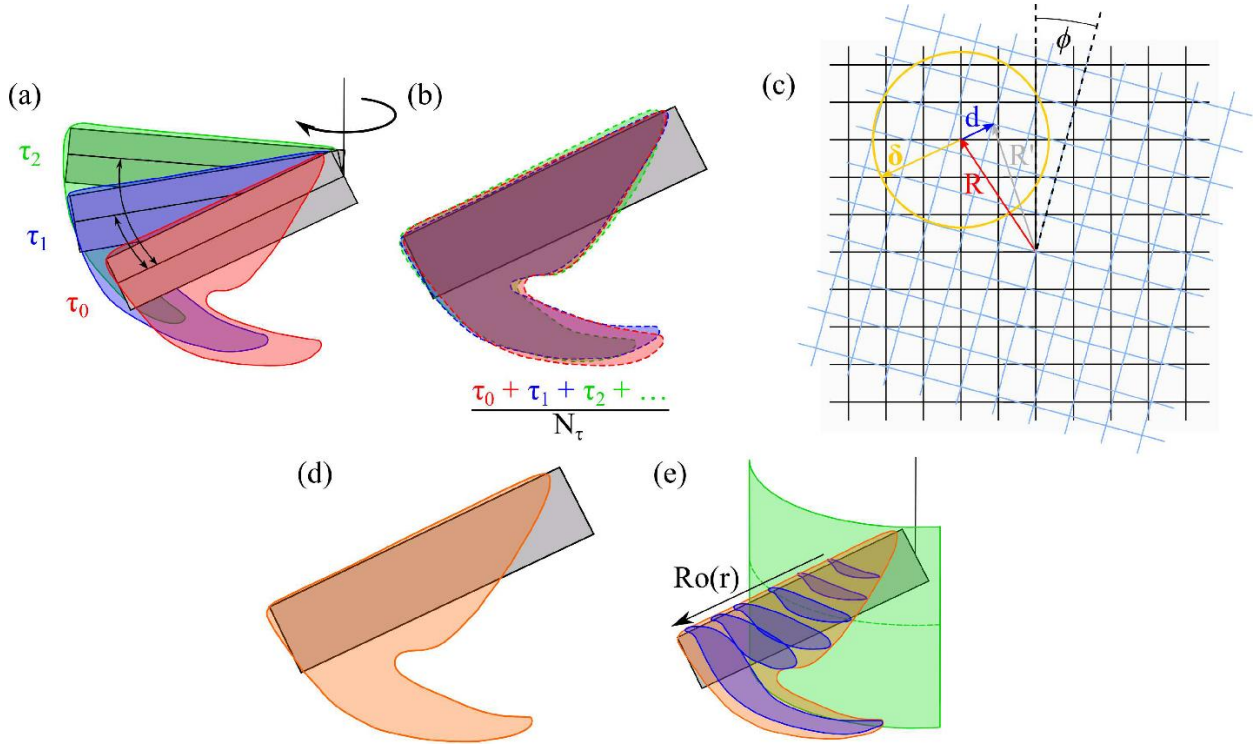

**Figure S1**. Calculating the time and spatial-averages. (a) The LEV control-volume moves forward with each increasing time step. (b) The LEV control-volumes are returned back to the initial angle of the first time step $\tau_0$. The data within each of these LEV control-volumes are averaged for $N_\tau$ number of time steps. (c) For each rotated LEC control-volume the original mesh and rotated mesh are misaligned by an angle $\phi$. The meshes are realigned into a single mesh by averaging all the points a distance $\delta = 0.8\sqrt{(\Delta x)^2 + (\Delta z)^2}$ and assigning the new value at the location $R$. (d) The time-averaged LEV control-volume is the combination of all the individual control-volumes from each time step. (e) The area-averages are calculated using the points within a small tolerance of the intersection between the cylindrical surfaces (green) corresponding to different $Ro(r)$ and the time-averaged LEV control-volume. The volume-averages are a weighted average of the area-averages using the number of slices corresponding to the $Ro(r)$.

## Supplementary Material

This section elaborates on the time and spatial-averaging process discussed in §2.3 using the data from the $AR = 5$ simulations. The time-averaging process considers the LEV control-volume at each time step. This is illustrated for three individual time steps (\tau_0, \tau_1, \tau_2) in the third revolution S1a. Since the leading-edge moves forward at each time-step, the region of the flow within the LEV control-volume changes. The spatial-averaging considers surfaces coinciding with cylinders with radii at different local Rossby numbers $Ro(r)$ bounded by the time-averaged LEV control-volume. This is shown in figure S1e.

Each of the time-averaged terms from equation 2.6 are displayed in figures S3 and S4 for Re=110 and 1400 respectively, with the same symbols and colors as are is in figure 6. Additionally, the time-averaged and normalized time derivative of radial vorticity $\widehat{W}^*$ is plotted using pink stars

N. Werner, H. Chung, J. Wang, G. Liu, J. Cimbala, H. Dong, and B. Cheng

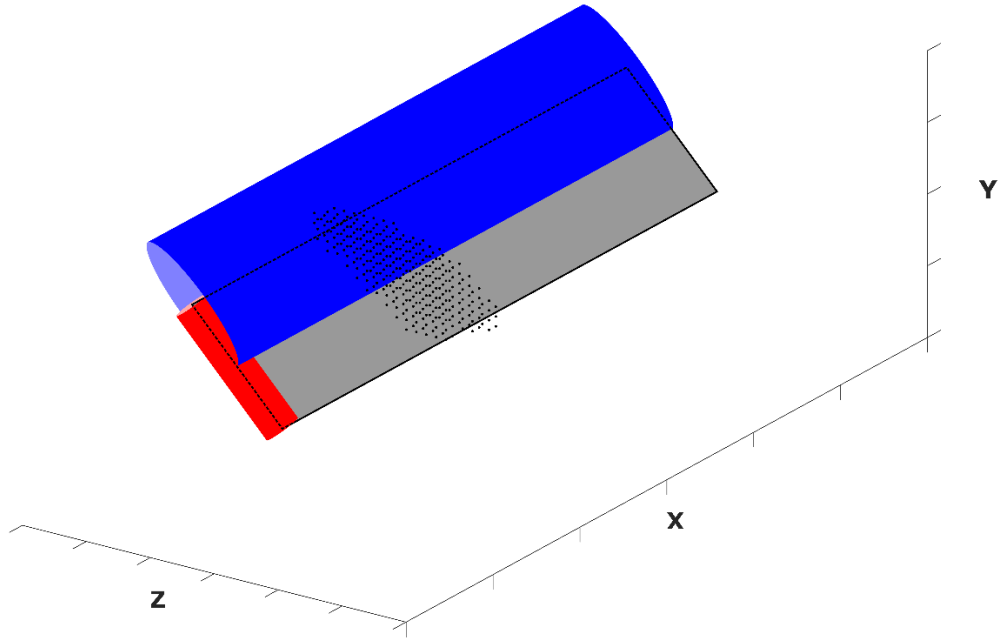

**Figure S2**. Both the leading-edge cylinder (LEC) and tip-cylinder (TC) are designed as elliptical. Therefore each cylinder will be designated by a major radius $k$ and minor radius $h$. The radii for the LEC are designated by $k_1$ and $h_1$, while the radii for the TC are $k_2$ and $h_2$ respectively. If there is no cylinder used in the figure then both $k = h = 0$. The LEC is shown in blue with radii $k_1 = 0.5c$ and $h_1 = 0.0625c$, and the TC in red with radii $k_2 = h_2 = 0.0625c$. The black dots represent the data points inside the control-volume defined by the time-averaged radial vorticity $\overline{\omega_r^*} = -3$.

along with the other terms. In figures S3a and S4a, neither the LEC or TC are used and the $\widehat{W}^*$ term is large and non-zero for the majority of the wingspan. In figures S3b and S4b, only the TC is applied, while in figures S3c and S4c only the LEC is used.

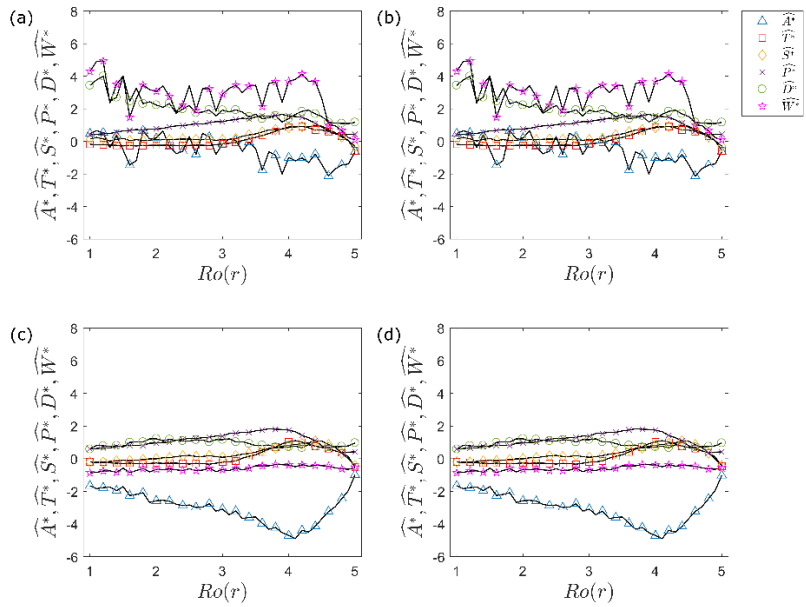

**Figure S3**. The effect of using the leading-edge-cylinder (LEC) and tip-cylinder (TC) at $AR = 5$ and $Re = 110$ on the local vorticity dynamics results. (a) $k_1 = h_1 = 0$ and $k_2 = h_2 = 0$. (b) $k_1 = h_1 = 0$, and $k_2 = h_2 = 0.0625c$. (c) $k_1 = 0.5c$, $h_1 = 0.0625c$, and $k_2 = h_2 = 0$. (d) $k_1 = 0.5c$, $h_1 = 0.0625c$, and $k_2 = h_2 = 0.0625c$.

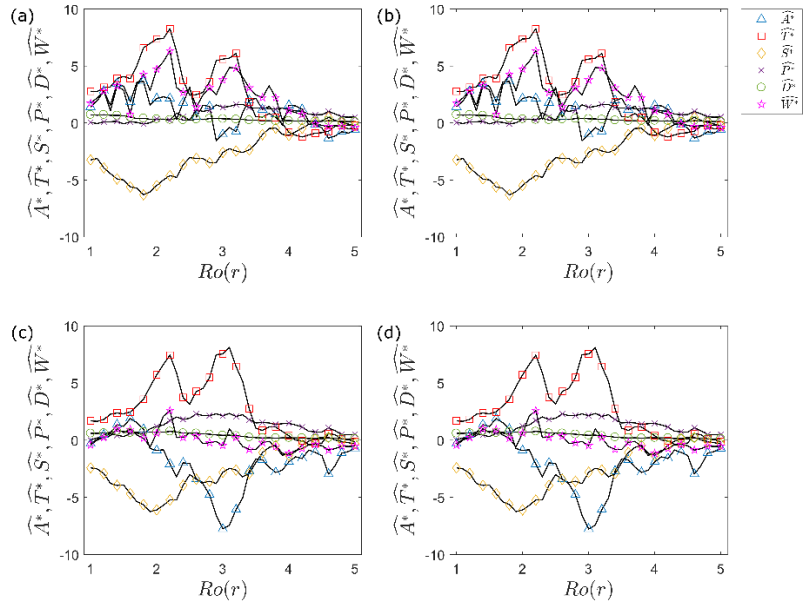

**Figure S4**. The effect of using the leading-edge-cylinder (LEC) and tip-cylinder (TC) at $AR = 5$ and $Re = 1400$ on the local vorticity dynamics results. (a) $k_1 = h_1 = 0$ and $k_2 = h_2 = 0$. (b) $k_1 = h_1 = 0$, and $k_2 = h_2 = 0.0625c$. (c) $k_1 = 0.5c$, $h_1 = 0.0625c$, and $k_2 = h_2 = 0$. (d) $k_1 = 0.5c$, $h_1 = 0.0625c$, and $k_2 = h_2 = 0.0625c$.